
\newif\ifproofmode			
\proofmodefalse				

\newif\ifforwardreference		
\forwardreferencetrue			

\newif\ifeqchapternumbers		
\eqchapternumbersfalse			

\newif\ifsectionnumbers			
\sectionnumberstrue			

\newif\ifeqsectionnumbers		
\eqsectionnumbersfalse			

\newif\ifchaptersectionnumbers     	
\chaptersectionnumberstrue		

\newif\ifcontinuoussectionnumbers	
\continuoussectionnumbersfalse	

\newif\ifcontinuousnumbers		
\continuousnumbersfalse 		

\newif\iffigurechapternumbers		
\figurechapternumbersfalse		

\newif\ifcontinuousfigurenumbers	
\continuousfigurenumbersfalse		

\newif\ifcontinuousreferencenumbers     
\continuousreferencenumberstrue         

\newif\ifparenequations			
\parenequationstrue			

\newif\ifstillreading			

\font\eqsixrm=cmr6			
\def\marginstyle{\eqsixrm}		

\newtoks\chapletter			
\newcount\chapno			
\newcount\sectno			
\newcount\eqlabelno			
\newcount\figureno			
\newcount\referenceno			
\newcount\minutes			
\newcount\hours				

\newread\labelfile			
\newwrite\labelfileout			
\newwrite\allcrossfile			

\chapno=0
\sectno=0
\eqlabelno=0
\figureno=0


\def\chapternumberstrue{\eqchapternumberstrue}

%
\def\initialeqmacro{
    \ifproofmode
        \headline{\tenrm \today\ --\ \timeofday\hfill
                         \jobname\ --- draft\hfill\folio}
        \hoffset=-1cm
        \immediate\openout\allcrossfile=zallcrossreferfile
    \fi
    \ifforwardreference
        \openin\labelfile=zlabelfile
        \ifeof\labelfile
        \else
            \stillreadingtrue
            \loop
                \read\labelfile to \nextline
                \ifeof\labelfile
                    \stillreadingfalse
                \else
                    \nextline
                \fi
                \ifstillreading
            \repeat
        \fi
        \immediate\openout\labelfileout=zlabelfile
    \fi}


{\catcode`\^^I=9
\catcode`\ =9
\catcode`\^^M=9
\endlinechar=-1
\globaldefs=1


%
\def\chapfolio{			
    \ifnum \chapno>0 \relax
        \the\chapno
    \else
        \the\chapletter
    \fi}

%
\def\bumpchapno{
    \ifnum \chapno>-1 \relax
        \global \advance \chapno by 1
    \else
        \global \advance \chapno by -1 \setletter\chapno
    \fi
    \ifcontinuousnumbers
    \else
        \global\eqlabelno=0
    \fi
    \ifcontinuousfigurenumbers
    \else
        \global\figureno=0
    \fi
    \ifcontinuousreferencenumbers
    \else
        \global\referenceno=0
    \fi
    \sectno=0}

\def\bumpsectno{
    \global\advance\sectno by 1 \relax
    \ifeqsectionnumbers
        \ifcontinuoussectionnumbers
        \else
            \global\eqlabelno=0
        \fi
    \fi}

%
\def\setletter#1{\ifcase-#1 {}  \or\global\chapletter={A}
  \or\global\chapletter={B} \or\global\chapletter={C} \or\global\chapletter={D}
  \or\global\chapletter={E} \or\global\chapletter={F} \or\global\chapletter={G}
  \or\global\chapletter={H} \or\global\chapletter={I} \or\global\chapletter={J}
  \or\global\chapletter={K} \or\global\chapletter={L} \or\global\chapletter={M}
  \or\global\chapletter={N} \or\global\chapletter={O} \or\global\chapletter={P}
  \or\global\chapletter={Q} \or\global\chapletter={R} \or\global\chapletter={S}
  \or\global\chapletter={T} \or\global\chapletter={U} \or\global\chapletter={V}
  \or\global\chapletter={W} \or\global\chapletter={X} \or\global\chapletter={Y}
  \or\global\chapletter={Z}\fi}

%
\def\tempsetletter#1{\ifcase-#1 {}\or{} \or\chapletter={A} \or\chapletter={B}
 \or\chapletter={C} \or\chapletter={D} \or\chapletter={E}
  \or\chapletter={F} \or\chapletter={G} \or\chapletter={H}
   \or\chapletter={I} \or\chapletter={J} \or\chapletter={K}
    \or\chapletter={L} \or\chapletter={M} \or\chapletter={N}
     \or\chapletter={O} \or\chapletter={P} \or\chapletter={Q}
      \or\chapletter={R} \or\chapletter={S} \or\chapletter={T}
       \or\chapletter={U} \or\chapletter={V} \or\chapletter={W}
        \or\chapletter={X} \or\chapletter={Y} \or\chapletter={Z}\fi}

%
\def\chapshow#1{
    \ifnum #1>0 \relax
        #1
    \else
        {\tempsetletter{\number#1}\the\chapletter}
    \fi}

%
\def\today{\number\day\space \ifcase\month\or Jan\or Feb\or
        Mar\or Apr\or May\or Jun\or Jul\or Aug\or Sep\or
        Oct\or Nov\or Dec\fi, \space\number\year}

\def\timeofday{\minutes=\time    \hours=\time
        \divide \hours by 60
        \multiply \hours by 60
        \advance \minutes by -\hours
        \divide \hours by 60
        \ifnum\the\minutes>9 \relax
     		\the\hours:\the\minutes
 	\else
  		\the\hours:0\the\minutes
	\fi}


%
%
%
%
\def\chapnum{\bumpchapno \chapfolio}

\def\chaplabel#1{
    \ifforwardreference                             
        \write\labelfileout{                        
        \noexpand\expandafter\noexpand\def          
        \noexpand\csname CHAPLABEL#1\endcsname{\the\chapno}}
    \fi
    \global\expandafter\edef\csname CHAPLABEL#1\endcsname
    {\the\chapno}
    \ifproofmode
        \rlap{\hbox{\marginstyle #1\ }}
    \fi}

%
\def\sectnum{
    \bumpsectno
        \ifchaptersectionnumbers
            \chapfolio.
        \fi
    \the\sectno}

\def\sectlabel#1{
    \bumpsectno
    \ifforwardreference
        \immediate\write\labelfileout{
        \noexpand\expandafter\noexpand\def
        \noexpand\csname SECTLABEL#1\endcsname{\the\chapno.\the\sectno?!}}
    \fi
    \global\expandafter\edef\csname SECTLABEL#1\endcsname
    {\the\chapno.\the\sectno?!}	 			
    \ifproofmode
        \llap{\hbox{\marginstyle #1\ }}
    \fi
    \ifchaptersectionnumbers
        \chapfolio.
    \fi
    \the\sectno}

\def\sectref#1{                                  
    \ifundefined{SECTLABEL#1}                     
        ++                                        
        \ifproofmode
            \ifforwardreference
            \else
                \write16{ ***Undefined\space Section\space Reference #1*** }
            \fi
        \else
            \write16{ ***Undefined\space Section\space Reference #1*** }
        \fi
    \else
        \edef\LABxx{\getlabel{SECTLABEL#1}}
	\ifchaptersectionnumbers
            \def\LAByy{\expandafter\stripchap\LABxx}
	    \chapshow\LAByy.
	\fi
	\expandafter\stripsect\LABxx
    \fi
    \ifproofmode
        \write\allcrossfile{Section\space #1}
    \fi}

%
%
\def\eqnum{                                    
    \global\advance\eqlabelno by 1              
    \eqno(
    \ifeqchapternumbers
        \chapfolio.
    \fi
    \ifeqsectionnumbers
        \the\sectno.
    \fi
    \the\eqlabelno)}

\def\eqlabel#1{                                
    \global\advance\eqlabelno by 1              
    \ifforwardreference                     
        \immediate\write\labelfileout{\noexpand\expandafter\noexpand\def
        \noexpand\csname EQLABEL#1\endcsname
        {\the\chapno.\the\sectno?\the\eqlabelno!}}
    \fi
    \global\expandafter\edef\csname EQLABEL#1\endcsname
    {\the\chapno.\the\sectno?\the\eqlabelno!}
    \eqno(
    \ifeqchapternumbers
        \chapfolio.
    \fi
    \ifeqsectionnumbers
        \the\sectno.
    \fi
    \the\eqlabelno)
    \ifproofmode
        \rlap{\hbox{\marginstyle #1}}		
    \fi}

\def\eqalignnum{                               
    \global\advance\eqlabelno by 1              
    &(\ifeqchapternumbers
        \chapfolio.
    \fi
    \ifeqsectionnumbers
        \the\sectno.
    \fi
    \the\eqlabelno)}

\def\eqalignlabel#1{                   	
    \global\advance\eqlabelno by 1 	        
    \ifforwardreference                     
        \immediate\write\labelfileout{\noexpand\expandafter\noexpand\def
        \noexpand\csname EQLABEL#1\endcsname
        {\the\chapno.\the\sectno?\the\eqlabelno!}}
    \fi
    \global\expandafter\edef\csname EQLABEL#1\endcsname
    {\the\chapno.\the\sectno?\the\eqlabelno!}
    &(\ifeqchapternumbers
        \chapfolio.
    \fi
    \ifeqsectionnumbers
        \the\sectno.
    \fi
    \the\eqlabelno)
    \ifproofmode
        \rlap{\hbox{\marginstyle #1}}			
    \fi}

\def\dnum{                                     
    \global\advance\eqlabelno by 1              
    \llap{(	 				
    \ifeqchapternumbers
        \chapfolio.
    \fi
    \ifeqsectionnumbers
        \the\sectno.
    \fi
    \the\eqlabelno)}}

\def\dlabel#1{                                 
    \global\advance\eqlabelno by 1              
    \ifforwardreference                         
        \immediate\write\labelfileout{\noexpand\expandafter\noexpand\def
        \noexpand\csname EQLABEL#1\endcsname
        {\the\chapno.\the\sectno?\the\eqlabelno!}}
    \fi
    \global\expandafter\edef\csname EQLABEL#1\endcsname
    {\the\chapno.\the\sectno?\the\eqlabelno!}
    \llap{(
    \ifeqchapternumbers
        \chapfolio.
    \fi
    \ifeqsectionnumbers
        \the\sectno.
    \fi
    \the\eqlabelno)}
    \ifproofmode
        \rlap{\hbox{\marginstyle #1}}		
    \fi}

\def\eqref#1{\ifparenequations(\fi
    \ifundefined{EQLABEL#1}***
        \ifproofmode
            \ifforwardreference
            \else
                \write16{ ***Undefined\space Equation\space Reference #1*** }
            \fi
        \else
            \write16{ ***Undefined\space Equation\space Reference #1*** }
        \fi
    \else
        \edef\LABxx{\getlabel{EQLABEL#1}}
	\def\LAByy{\expandafter\stripsect\LABxx}
        \def\LABzz{\expandafter\stripchap\LABxx}
        \ifeqchapternumbers
            \chapshow{\LABzz}.
        \else
            \ifnum \number\LABzz=\chapno \relax
            \else
                \chapshow{\LABzz}.
            \fi
        \fi
	\ifeqsectionnumbers
	    \LAByy.
	\fi
        \expandafter\stripeq\LABxx
    \fi
    \ifparenequations)\fi
    \ifproofmode
        \write\allcrossfile{Equation\space #1}
    \fi}

%
\def\fignum{                                   
    \global\advance\figureno by 1\relax         
    \iffigurechapternumbers
        \chapfolio.
    \fi
    \the\figureno}

\def\figlabel#1{				
    \global\advance\figureno by 1\relax 	
    \ifforwardreference				
        \immediate\write\labelfileout{\noexpand\expandafter\noexpand\def
        \noexpand\csname FIGLABEL#1\endcsname
        {\the\chapno.\the\sectno?\the\figureno!}}
    \fi
    \global\expandafter\edef\csname FIGLABEL#1\endcsname
    {\the\chapno.\the\sectno?\the\figureno!}
    \iffigurechapternumbers
        \chapfolio.
    \fi
    \ifproofmode
        \llap{\hbox{\marginstyle #1\ }}\relax
    \fi
    \the\figureno}

\def\figref#1{					
    \ifundefined				
        {FIGLABEL#1}!!!!			
        \ifproofmode
            \ifforwardreference
            \else
                \write16{ ***Undefined\space Figure\space Reference #1*** }
            \fi
        \else
            \write16{ ***Undefined\space Figure\space Reference #1*** }
        \fi
    \else
        \edef\LABxx{\getlabel{FIGLABEL#1}}
        \def\LABzz{\expandafter\stripchap\LABxx}
        \iffigurechapternumbers
            \chapshow{\LABzz}.\expandafter\stripeq\LABxx
        \else \ifnum\number\LABzz=\chapno \relax
                \expandafter\stripeq\LABxx
            \else
                \chapshow{\LABzz}.\expandafter\stripeq\LABxx
            \fi
        \fi
        \ifproofmode
            \write\allcrossfile{Figure\space #1}
        \fi
    \fi}

%
%
\def\pagelabel#1{
    \ifforwardreference
        \write\labelfileout{
        \noexpand\expandafter\noexpand\def
        \noexpand\csname PGLABEL#1\noexpand\endcsname{\the\pageno}}
    \fi
    \global\expandafter\edef\csname PGLABEL#1\endcsname{\the\pageno}}

\def\pageref#1{
    \ifundefined
        {PGLABEL#1}***
        \ifproofmode
        \else
            \write16{ ***Undefined\space Page\space Reference #1*** }
        \fi
    \else
        \csname PGLABEL#1\endcsname
    \fi
    \ifproofmode
        \write\allcrossfile{Page\space #1}
    \fi}

%
\def\refnum{                                      
    \global\advance\referenceno by 1\relax         
    \the\referenceno}	                           

\def\internalreflabel#1{			
    \global\advance\referenceno by 1\relax 	
    \ifforwardreference				
        \immediate\write\labelfileout{\noexpand\expandafter\noexpand\def
        \noexpand\csname REFLABEL#1\endcsname
        {\the\chapno.\the\sectno?\the\referenceno!}}
    \fi
    \global\expandafter\edef\csname REFLABEL#1\endcsname
    {\the\chapno.\the\sectno?\the\figureno!}
    \ifproofmode
        \llap{\hbox{\marginstyle #1\hskip.5cm}}\relax
    \fi
    \the\referenceno}

\def\internalrefref#1{				
    \ifundefined				
        {REFLABEL#1}!!!!			
        \ifproofmode
            \ifforwardreference
            \else
                \write16{ ***Undefined\space Footnote\space Reference #1*** }
            \fi
        \else
            \write16{ ***Undefined\space Footnote\space Reference #1*** }
        \fi
    \else
        \edef\LABxx{\getlabel{REFLABEL#1}}
        \def\LABzz{\expandafter\stripchap\LABxx}
        \expandafter\stripeq\LABxx
        \ifproofmode
            \write\allcrossfile{Reference\space #1}
        \fi
    \fi}

%
\def\reflabel#1{\item{\internalreflabel{#1}.}}

%
\def\refref#1{\internalrefref{#1}}

\def\eq{\ifhmode Eq.~\else Equation~\fi}		
\def\eqs{\ifhmode Eqs.~\else Equations~\fi}

%
%
%
%

%
\def\getlabel#1{\csname#1\endcsname}
\def\ifundefined#1{\expandafter\ifx\csname#1\endcsname\relax}
\def\stripchap#1.#2?#3!{#1}			
\def\stripsect#1.#2?#3!{#2}			%
\def\stripeq#1.#2?#3!{#3}			
}  

\overfullrule = 0pt
\magnification = \magstep1
\baselineskip 14pt
\hsize = 6.0 truein
\vsize = 8.5 truein
\chapternumberstrue
\forwardreferencetrue
\initialeqmacro

\def\ch{\mathop{\rm ch}\nolimits}
\def\cth{\mathop{\rm cth}\nolimits}
\def\tr{\mathop{\rm tr}\nolimits}

\line{\hfill UMTG-170}


\vglue 0.4 truein

\centerline{{\bf INTRODUCTION TO THE THERMODYNAMICS OF SPIN CHAINS}
\footnote*{To appear in the Proceedings of the NSERC-CAP Workshop on Quantum
Groups, Integrable Models and Statistical Systems, Kingston, Canada
13 -17 July 1992}}
\bigskip

\medskip

\centerline{LUCA MEZINCESCU and RAFAEL I. NEPOMECHIE}
\centerline{Department of Physics}
\centerline{University of Miami, Coral Gables, FL 33124, USA}

\vskip 0.2 in

\bigskip

\centerline{\bf ABSTRACT}

\vskip 0.2 in

{\leftskip .5truein \rightskip .5truein  \baselineskip 12pt
We review how to obtain the thermodynamic Bethe Ansatz (TBA) equations for the
antiferromagnetic Heisenberg ring in an external magnetic field. We review how
to solve these equations for low temperature and small field, and calculate the
specific heat and magnetic susceptibility. \par}

\vskip 0.2 in

\bigskip

\medskip

\noindent
{\bf \chapnum . Motivation}
\vskip 0.2truein

Quantum spin chains provide some of the simplest examples of physical
systems with quantum-algebra${}^{\refref{qsymmetry}}$ symmetry.
A well-known example is the open quantum spin chain consisting
of $N$ spins with the Hamiltonian
$$\eqalignno{
{\cal H} &= \sum_{k=1}^{N-1} \Bigl\{ \sigma^1_k \sigma^1_{k+1} +
\sigma^2_k \sigma^2_{k+1} + {1\over 2}( q + q^{-1}) \sigma^3_k \sigma^3_{k+1}
\Bigr\} \cr
&\quad\quad\quad\quad\quad - {1\over 2}( q - q^{-1})
\Bigl( \sigma^3_1 - \sigma^3_N \Bigr) \,, \eqalignlabel{xxz} \cr}
$$
where $\vec\sigma$ are the usual Pauli matrices, and $q$ is an arbitrary
complex parameter. This model is${}^{\refref{alcaraz}, \refref{sklyanin}}$
integrable and the Hamiltonian
commutes${}^{\refref{pasquier},\refref{kulish/sklyanin(1)}}$ with the
generators of the quantum algebra $U_q[su(2)]$.

This model can be generalized in various directions. We have considered
integrable open chains with quantum-algebra symmetry that have instead spins in
higher-dimensional representations${}^{\refref{spin1}}$, or spins in the
fundamental representation of larger symmetry algebras${}^{\refref{general}}$.
(See also Ref. \refref{batchelor}.)

We would like to determine the physical properties (e.g., low-temperature
magnetic susceptibility and specific heat) of such models for $|q| =1$ in
the thermodynamic ($N \rightarrow \infty$) limit. For the model \eqref{xxz},
some steps in this direction have already been taken in Ref. \refref{pasquier}.
Because these models are integrable, the spectra of their Hamiltonians are
known implicitly through the Bethe Ansatz (BA) equations. Unfortunately, this
is still quite far from determining the models' physical properties.

We consider here instead, as a warm-up exercise, the analogous problem
for the antiferromagnetic Heisenberg ring, with Hamiltonian
$${\cal H} = {1\over 4}\sum_{k=1}^{N} \left( \vec \sigma_k \cdot
\vec \sigma_{k+1} - 1 \right) - {H \over 2} \sum_{k=1}^{N} \sigma^3_k
\,, \qquad\qquad \vec \sigma_{N+1} = \vec \sigma_1 \,,
\eqlabel{xxx}
$$
where $H ( \ge 0)$ is an external magnetic field. Although this model is
comparatively simpler, the methods used to investigate it can presumably
also be used to study integrable chains with quantum-algebra symmetry.

The low-temperature thermodynamics of this model was determined approximately
twenty years ago as the result of the work of many people:
Yang and Yang${}^{\refref{yangyang1}}$, Takahashi${}^{\refref{takahashi1}}$,
Gaudin${}^{\refref{gaudin}}$, Johnson and McCoy${}^{\refref{johnson/mccoy}}$,
{\it et al}. The fact that this work (certain details of which have never been
published) is scattered among many papers contributed to the difficulty we
experienced in learning about it.

We summarize this work here, in the hope of making it more accessible to
others. Following Takahashi, we formulate in Section 2 the
so-called thermodynamic Bethe Ansatz (TBA) equations for the
antiferromagnetic Heisenberg ring. These are an infinite set of coupled
nonlinear integral equations, which describe the equilibrium
thermodynamics of the model at temperature $T$ and in the field $H$. Following
Johnson and McCoy, in Section 3 we solve the TBA equations in a systematic
low-$T$ and small-$H$ perturbative expansion; and we calculate the free energy,
specific heat, and magnetic susceptibility.

We warn the reader that we shall encounter along the way several
arguments that are far from rigorous. Nevertheless, as far as we know, this is
still the present state of the art.

While there are other ways${}^{\refref{takahashi2}, \refref{filyov}}$
of calculating the low-temperature specific heat within the general TBA
approach, the method pursued here has the merit of treating this calculation
in the same manner as the one for the magnetic susceptibility.
Review articles on the thermodynamics of integrable models which we have found
particularly useful are Refs. \refref{faddeev1} - \refref{bogoliubov}.

\vskip 0.4truein
\noindent
{\bf \chapnum .  TBA equations}

\vskip 0.2truein

The Hamiltonian for the antiferromagnetic Heisenberg ring in a magnetic field
is given by \eq\eqref{xxx}. The corresponding energy eigenvalues, which
can be determined by either the coordinate${}^{\refref{bethe}}$,
algebraic${}^{\refref{faddeev2}}$, or analytic${}^{\refref{reshetikhin}}$
Bethe Ansatz, are given by
$$E = - {1\over 2}\sum_{\alpha=1}^M {1 \over \lambda^2_\alpha + {1\over 4}}
 - H \left( {N\over 2} - M \right)
\,, \eqlabel{eigenvalues} $$
where the variables $\{ \lambda_\alpha \}$ satisfy the Bethe Ansatz (BA)
equations
$$\eqalignno{
\left({\lambda_\alpha + {i\over 2} \over
       \lambda_\alpha - {i\over 2}} \right)^N
= -& \prod_{\beta=1}^M {\lambda_\alpha - \lambda_\beta + i \over
                        \lambda_\alpha - \lambda_\beta - i} \,,
\qquad \alpha = 1, \cdots , M \,, \qquad M \le {N\over 2} \,.
\eqalignlabel{BA} \cr} $$
These equations can be written more compactly as
$$ \left[ e_1(\lambda_\alpha)\right]^N = - \prod_{\beta=1}^M
e_2(\lambda_\alpha - \lambda_\beta) \,, \eqlabel{BAcompact} $$
where
$$ e_n (\lambda) \equiv {\lambda + {in\over 2}\over \lambda - {in\over 2}}
\,. \eqnum $$

These equations are difficult to solve for finite $N$. Fortunately, in the
physically-interesting thermodynamic ($N \rightarrow \infty$) limit,
these equations become easier to solve. Indeed, one then finds (see, e.g.,
Ref. \refref{faddeev1}) that the BA equations have ``string'' solutions.
For instance, a string of length 1 is
$$ \lambda^{(1)} = x \,, $$
where $x$ is an arbitrary real number, which is called the ``center'' of the
string. Moreover, a string of length 2 is
$$ \left\{ \matrix{\lambda^{(1)} = x + {i\over 2} \cr
                   \lambda^{(2)} = x - {i\over 2} \cr} \right. \,, $$
a string of length 3 is
$$ \left\{ \eqalign{\lambda^{(1)} &= x + i \cr
                    \lambda^{(2)} &= x     \cr
                    \lambda^{(3)} &= x - i \cr} \right. \,, $$
and so on. We adopt the ``string hypothesis'' which states that {\it all} the
solutions $\{ \lambda_\alpha \,, \alpha = 1, \cdots , M \}$ organize
themselves into such strings. That is, the solutions are collections
of $M_n$ strings of length $n$ of the form
$$\lambda_\alpha^{(n,j)} = \lambda_\alpha^n + i \left( {n+1\over 2} - j \right)
\,, \eqlabel{string} $$
where $j = 1, \cdots , n$; $\alpha = 0,1, \cdots, M_n$;
$n = 1, \cdots, \infty$; and the centers $\lambda_\alpha^n$ are real.
A particular solution of the
BA equations corresponds to a set of non-negative integers $\{M_n \}$
and the $M_n$ real numbers $\lambda_\alpha^n$ for each $n$. Observe that
the total number of $\lambda$ variables, and BA equations, is
$M = \sum_{n=1}^\infty n M_n$.

Under this hypothesis, the BA equations \eqref{BAcompact} become
$$ \left[ e_1(\lambda_\alpha^{(n,j)})\right]^N = - \prod_{m=1}^\infty
\prod_{\beta=0}^{M_m} \prod_{k=1}^m
e_2(\lambda_\alpha^{(n,j)} - \lambda_\beta^{(m,k)}) \,. \eqnum $$
Following Takahashi${}^{\refref{takahashi1}}$ and Gaudin${}^{\refref{gaudin}}$,
we form the product $\prod_{j=1}^n$, and obtain a set of equations for
the ({\it real}) centers $\lambda_\alpha^n$:
$$ \left[ e_n(\lambda_\alpha^n)\right]^N = (-)^n \prod_{m=1}^\infty
\prod_{\beta=0}^{M_m}
E_{nm}(\lambda_\alpha^n - \lambda_\beta^m) \,, \eqlabel{BAcenters} $$
where
$$E_{nm}(\lambda) \equiv e_{|n-m|}(\lambda)\ e_{|n-m|+2}^2(\lambda)\ \cdots
\ e_{n+m-2}^2(\lambda)\ e_{n+m}(\lambda) \,.
\eqnum $$
In obtaining this result, we use the lemmas
$$ \prod_{j=1}^n e_m (\lambda_\alpha^{(n,j)}) = \prod_{l=1}^{min(n,m)}
e_{n+m+1-2l} (\lambda_\alpha^n) \,, \eqlabel{lemma1} $$
$$ \prod_{j=1}^n \prod_{k=1}^m  e_2 (\lambda_\alpha^{(n,j)}
- \lambda_\beta^{(m,k)}) = E_{nm} (\lambda_\alpha^n -\lambda_\beta^m) \,.
\eqlabel{lemma2} $$

Since the equations \eqref{BAcenters} involve only products of phases,
it is useful to take the logarithm:
$$N p_n(\lambda_\alpha^n) = 2\pi J_\alpha^n
+ \sum_{m=1}^\infty \sum_{\beta=0}^{M_{m}}  \Xi_{n m}
(\lambda_\alpha^n - \lambda_\beta^{m})  \,, \eqlabel{BAlog} $$
where
$$
p_n(\lambda) \equiv i \ln e_n (\lambda) \,, \qquad\qquad
\Xi_{n m}(\lambda) \equiv i \ln E_{n m} (\lambda) \,, \eqnum $$
and $J_\alpha^n$ are integers or half-integers constrained by
$$ -J_{max}^n \le J_\alpha^n \le J_{max}^n  \,.  \eqlabel{range} $$
Faddeev and Takhtajan have given${}^{\refref{faddeev1}}$ a prescription
for determining an explicit expression for $J_{max}^n$ from \eqref{BAlog}.
Any integer or half-integer in the range \eqref{range} is called
``admissible''. We adopt the important hypothesis that the numbers
$\{ J_\alpha^n \}$ can be regarded as quantum numbers of the model: for every
admissible set $\{ J_\alpha^n \}$ (no two of which are identical),
there is a unique solution $\{ \lambda_\alpha^n \}$ (no two of which are
identical) of \eq\eqref{BAlog}. Such solutions are called ``particle
rapidities''.

As we shall see, it is useful to introduce (following Yang and
Yang${}^{\refref{yangyang1}}$) the concept of ``hole''. Given
a set of quantum numbers $\{ J_\alpha^n \}$, we define the set
of numbers $\{ \tilde J_\alpha^n \}$ to be the admissible values {\it not}
in $\{ J_\alpha^n \}$. We define corresponding quantities
$\{ \tilde \lambda_\alpha^n \}$, called ``hole rapidities'', by
$$N p_n(\tilde \lambda_\alpha^n) = 2\pi \tilde J_\alpha^n
+ \sum_{m=1}^\infty \sum_{\beta=0}^{M_{m}}  \Xi_{n m}
(\tilde\lambda_\alpha^n - \lambda_\beta^{m})  \,. \eqlabel{holes} $$

For $N \rightarrow \infty$, the distributions of the numbers
$\{ \lambda_\alpha^n \}$ and $\{ \tilde\lambda_\alpha^n \}$ along the
real axis are characterized by corresponding particle and hole densities
$\rho_n (\lambda)$ and $\tilde\rho_n (\lambda)$, respectively:
$$\eqalignno{
N \rho_n (\lambda) d\lambda &= {\hbox{ number of   }} \lambda^n
{\hbox{'s in   }} d\lambda \cr
N \tilde \rho_n (\lambda) d\lambda &= {\hbox{ number of   }} \tilde\lambda^n
{\hbox{'s in   }} d\lambda  \,. \eqalignnum \cr} $$
We must now reformulate the description of the model in terms of these
densities.

We begin by exhibiting a constraint between the hole and particle densities.
To this end, we define the function
$$h^n (\lambda) \equiv {1\over 2\pi} \left\{ N p_n(\lambda)
-\sum_{m=1}^\infty \sum_{\beta=0}^{M_{m}}  \Xi_{n m}
(\lambda - \lambda_\beta^{m}) \right\} \,. \eqlabel{hn} $$
Evidently,
$$\eqalignno{
h^n (\lambda_\alpha^n) &= J_\alpha^n \,, \cr
h^n (\tilde\lambda_\alpha^n) &= \tilde J_\alpha^n \,. \eqalignnum \cr} $$
We make the assumption, which is implicit in Refs. \refref{takahashi1} and
\refref{gaudin}, that $h^n(\lambda)$ is a monotonic increasing function of
$\lambda$. It follows that
$$\eqalignno{
N \left[ \rho_n (\lambda) + \tilde\rho_n(\lambda) \right] d\lambda
& = {\hbox{ number of   }} \lambda^n {\hbox{'s and   }} \tilde\lambda^n
{\hbox{'s in   }} d\lambda \cr
& = {\hbox{ number of times   }} h^n(\lambda) {\hbox{   ranges over   }}
J_\alpha^n {\hbox{'s and   }} \tilde J_\alpha^n {\hbox{'s in   }} d\lambda \cr
& = h^n(\lambda + d\lambda) - h^n(\lambda) = d h^n \,. \eqalignnum \cr}
$$
That is,
$$ \rho_n (\lambda) + \tilde\rho_n(\lambda) = {1\over N} {dh^n \over d\lambda}
\,. \eqlabel{hprime} $$
Making in \eq\eqref{hn} the following replacement of sums by integrals
$$ {1\over N} \sum_{\beta = 0}^{M_m} \bigl( \ \ \bigr) \rightarrow
\int_{-\infty}^\infty \bigl( \ \ \bigr) \rho_m (\lambda') d \lambda' \,,
\eqnum $$
we see from \eqref{hprime} that
$$\rho_n(\lambda) + \tilde\rho_n(\lambda) =
{1\over 2\pi} {d p_n(\lambda)\over d\lambda}
-\sum_{m=1}^\infty \int_{-\infty}^\infty {1\over 2\pi}
{d \Xi_{n m}(\lambda - \lambda')\over d\lambda} \rho_m (\lambda') d \lambda'
\,. \eqnum $$
This can be rewritten in the final form
$$\tilde \rho_n +  \sum_{m=1}^\infty A_{nm} * \rho_{m} = a_n \,,
\eqlabel{constraint}$$
where
$$\eqalignno{
A_{nm}(\lambda) &\equiv {1\over 2\pi} {d \Xi_{n m}(\lambda)\over d\lambda}
+ \delta(\lambda) \delta_{nm} \cr
&= \delta(\lambda) \delta_{nm}
+ \left(1 - \delta_{nm} \right) a_{|n - m|}(\lambda) + a_{n + m}(\lambda) \cr
& \qquad\qquad +2\sum_{l=1}^{min(n,m)-1} a_{|n - m| + 2l}(\lambda) \,,
\eqalignlabel{Anm} \cr} $$
and
$$\eqalignno{
a_n(\lambda) &\equiv {1\over 2\pi} {d p_n (\lambda)\over d\lambda}
=  {1 \over 2\pi} {n\over \lambda^2 + {n^2\over 4}} \,,
\eqalignlabel{an} \cr} $$
and $*$ denotes the convolution
$$ \left( f * g \right) (\lambda) \equiv \int_{-\infty}^\infty
d\lambda'\ f(\lambda - \lambda') g(\lambda') = \left( g * f \right) (\lambda)
\,. \eqnum $$
One can regard the constraint equation \eqref{constraint} as a
definition of $\tilde \rho_n$ in terms of $\rho_n$.

We introduce here for future reference the ``inverse'' function
$$A^{-1}_{nm}(\lambda) \equiv \delta(\lambda) \delta_{nm}
- s(\lambda) \left( \delta_{n, m+1} + \delta_{n, m-1} \right) \,,
\eqlabel{inverse} $$
where $s(\lambda)$ is defined by
$$s(\lambda) \equiv {1\over 2 \ch \pi \lambda} \,. \eqlabel{s} $$
It has the properties that
$$\sum_{n'=1}^\infty \left( A^{-1}_{nn'} * A_{n' m} \right) (\lambda)
= \delta(\lambda) \delta_{nm}  \,, \eqnum $$
and
$$\sum_{m=1}^\infty \left( A^{-1}_{nm} * a_m \right) (\lambda)
=  s(\lambda) \delta_{n 1} \,, \qquad\qquad
\sum_{m=1}^\infty A^{-1}_{nm} * m = 0 \,. \eqlabel{Anmproperty} $$
These relations can be easily derived with the help of Fourier transforms,
for which we use the following conventions:
$$ \hat f(\omega) \equiv \int_{-\infty}^\infty e^{i \omega \lambda}\ f(\lambda)
d\lambda \,, \qquad\qquad
f(\lambda) = {1\over 2\pi} \int_{-\infty}^\infty e^{-i \omega \lambda}\
\hat f(\omega) d\omega \,. \eqnum $$
In particular, we note the convolution theorem
$$\int_{-\infty}^\infty e^{i \omega \lambda}\ \left( f * g \right)
(\lambda)\ d\lambda = \hat f(\omega)\ \hat g(\omega) \,, \eqnum $$
and the identities
$$ \hat a_n (\omega) = \left\{ \matrix{e^{-n |\omega|/2} & n > 0 \cr
                                       0                 & n = 0 \cr} \right.
\,, \qquad\qquad \hat s(\omega) = {1\over 2 \ch {\omega\over 2}} \,,
\eqnum $$
$$ \hat A_{nm} (\omega) =
\left( \cth {|\omega|\over 2} \right) \left[
e^{-|n - m| |\omega|/ 2} - e^{-(n + m)|\omega|/ 2} \right]
\,, \eqnum $$
$$ \hat A_{nm}^{-1} (\omega) =\delta_{nm} - \hat s(\omega) \left(
\delta_{n, m+1} + \delta_{n, m-1} \right) \,. \eqnum $$

We recall that the Bethe Ansatz provides the expression \eqref{eigenvalues}
for the energy eigenvalues in terms of $\lambda$'s. In order to recast
this expression in terms of densities, we first invoke the string hypothesis
to arrive at
$$E = - {1\over 2} \sum_{n=1}^\infty \sum_{\alpha=0}^{M_n} \sum_{j=1}^n
{1\over \left( \lambda_\alpha^{(n,j)} \right)^2 + {1\over 4}}
- H \left( {N\over 2} - \sum_{n=1}^\infty n M_n \right) \,. \eqnum $$
We next observe that
$$\eqalignno{
{1\over 2} \sum_{j=1}^n {1\over \left( \lambda_\alpha^{(n,j)} \right)^2
+ {1\over 4}} &= {i\over 2} {d\over d \lambda_\alpha^n} \sum_{j=1}^n
\ln e_1 (\lambda_\alpha^{(n,j)})
= {i\over 2} {d\over d \lambda_\alpha^n} \ln \prod_{j=1}^n
e_1 (\lambda_\alpha^{(n,j)}) \cr
&= {i\over 2} {d\over d \lambda_\alpha^n} \ln
e_n (\lambda_\alpha^n)
= {1\over 2} {d\over d \lambda_\alpha^n} p_n (\lambda_\alpha^n)
= \pi a_n (\lambda_\alpha^n) \,, \eqalignnum \cr} $$
where, in passing to the second line, we have used the lemma \eqref{lemma1}.
It follows that the energy is given by
$$\eqalignno{
E &= - \pi \sum_{n=1}^\infty \sum_{\alpha=0}^{M_n} a_n (\lambda_\alpha^n)
- N H \left( {1\over 2} - {1\over N} \sum_{n=1}^\infty n M_n \right) \cr
&= - N \pi \sum_{n=1}^\infty
\int_{-\infty}^\infty a_n(\lambda) \rho_n (\lambda) d\lambda
- N H \left( {1\over 2} - \sum_{n=1}^\infty n
\int_{-\infty}^\infty  \rho_n (\lambda) d\lambda \right)
\,. \eqalignlabel{energy} \cr} $$
Given a set of particle densities $\{ \rho_n \}$, this expression
determines an energy level of the system. Notice that this expression
does not depend on the hole densities $\tilde \rho_n$.

As an illustration of this formalism, let us briefly consider the
ground state of the system in zero field ($H=0$). We assume that
for this state, there are no holes. That is,
$$ \tilde \rho_n = 0 \,. \eqlabel{noholes} $$
(We shall prove this later from the TBA equations.) It follows from the
constraint \eqref{constraint} that
$$ \sum_{m=1}^\infty A_{nm} * \rho_{m} = a_n \,.  \eqnum$$
Thus, the particle densities are given by
$$ \rho_n = \sum_{m=1}^\infty A_{nm}^{-1} * a_m = s \delta_{n 1}
\,,  \eqlabel{rhozero} $$
where $s(\lambda)$ is given by \eqref{s}. That is, the antiferromagnetic
ground state is described by strings of length 1 ( i.e., real solutions of
the BA equations). From \eq\eqref{energy}, we conclude that the
energy per site for the ground state is
$$ e_0 = {E_0\over N} = - \pi \int_{-\infty}^\infty a_1(\lambda)
s (\lambda) d\lambda = - \ln 2 \,, \eqlabel{ezero} $$
which is the well-known result of Hulthen${}^{\refref{hulthen}}$.

This example raises two important questions: \hfill\break
\ (i) how to establish the result \eqref{noholes} that the ground state
has no holes? \hfill\break
(ii) how to study low-lying excitations? \hfill\break
An answer to both questions is to consider the system at temperature
$T \ge 0$. The equilibrium state at $T=0$ is the ground state;
and the equilibrium state at low but nonzero temperature provides information
about the low-lying excitations.

We now argue heuristically, following Refs. \refref{yangyang1},
\refref{andrei} and \refref{bogoliubov},
that the equilibrium state at temperature $T$ is described by certain
equilibrium densities $\rho_n^{eq}$ and $\tilde \rho_n^{eq}$. We begin
by observing that the partition function is given by
$$
Z = \tr e^{-{\cal H}/T} = \sum_{eigenstates} e^{-E/T}
= \sum_{\rho, \tilde \rho} \delta \left[ \chi (\rho, \tilde \rho) \right]\
W[\rho, \tilde \rho]\  e^{-E[\rho]/T} \,, \eqnum $$
where the delta function enforces the constraint
$$\chi (\rho, \tilde \rho) = 0 \,, \eqlabel{cc} $$
which represents the constraint \eqref{constraint}, and
$W[\rho, \tilde \rho]$ is equal to the number of states corresponding to
given densities $\rho$ and $\tilde \rho$. Introducing the entropy
$$ S = \ln W  \,,  \eqlabel{entropy} $$
and using the fact that the free energy is given by
$$F = E - T S \,, \eqlabel{free} $$
we see that
$$Z = \sum_{\rho, \tilde \rho}
\delta \left[ \chi (\rho, \tilde \rho) \right]\
e^{-\left( E[\rho] - T S[\rho, \tilde \rho] \right)/T}
= \sum_{\rho, \tilde \rho}
\delta \left[ \chi (\rho, \tilde \rho) \right]\
e^{-F [\rho, \tilde \rho]/T}
 \,. \eqnum $$
Using the saddle-point approximation (for $M \,, N \rightarrow \infty$
with $M/N =$ constant), we conclude that
$$Z = e^{-F[\rho_{eq}, \tilde \rho_{eq}]/T} \,, \eqnum $$
where the equilibrium densities are determined by the condition
$$ \delta F \Big\vert_{\rho = \rho_{eq}\,, \tilde \rho = \tilde \rho_{eq}}
= 0 \,, \eqnum $$
subject to the constraint \eqref{cc}.

In order to compute the entropy \eqref{entropy}, we observe (again, following
Yang and Yang${}^{\refref{yangyang1}}$) that
$$\eqalignno{
dS_n &= \ln \left\{ {\hbox{number of states (strings of length }}n
{\hbox{) in }}d\lambda \right\} \cr
&= \ln \left\{ {\hbox{number of ways of choosing }}\lambda^n
{\hbox{'s in }}d\lambda \right\} \cr
&= \ln \left\{ { \left[ N ( \rho_n + \tilde\rho_n) d\lambda \right]!
\over \left( N \rho_n d\lambda \right)!\
\left( N \tilde\rho_n d\lambda \right)!} \right\} \cr
&\approx N \left\{
\left( \rho_n + \tilde \rho_n \right) \ln \left( \rho_n + \tilde \rho_n \right)
- \rho_n \ln  \rho_n - \tilde\rho_n \ln  \tilde\rho_n  \right\} d\lambda
\,,  \eqalignnum \cr} $$
where, in passing to the last line, we have used Stirling's approximation.
We conclude that the entropy is given by
$$ S = \sum_{n=1}^\infty \int_{-\infty}^\infty dS_n
     = N \sum_{n=1}^\infty \int_{-\infty}^\infty
\left\{
\left( \rho_n + \tilde \rho_n \right) \ln \left( \rho_n + \tilde \rho_n \right)
- \rho_n \ln  \rho_n - \tilde\rho_n \ln  \tilde\rho_n  \right\} d\lambda
\,. \eqlabel{entropyfinal} $$
The hole concept was introduced precisely for calculating the entropy.

The free energy $F$ of the Heisenberg ring is given by \eqref{free},
with the energy $E$ given by \eqref{energy} and the entropy $S$ given
by \eqref{entropyfinal}. The condition $\delta F = 0$ subject to
the constraint \eqref{constraint} yields the thermodynamic Bethe
Ansatz (TBA) equations
$$T \ln \left( 1 + e^{\epsilon_{n}/T} \right)
= \sum_{m=1}^\infty A_{nm} * T \ln \left( 1 + e^{-\epsilon_m/T} \right)
-\pi a_n + n H \,, \eqlabel{TBA} $$
where the functions $\epsilon_n (\lambda)$ are defined by
$$\epsilon_n (\lambda) = T \ln \left({\tilde\rho_n^{eq} (\lambda)\over
\rho_n^{eq} (\lambda)}\right)\,.  \eqlabel{epsilon} $$
The TBA equations are an infinite set of coupled, nonlinear integral
equations, which we shall assume to have a unique solution
$\{ \epsilon_n (\lambda) \}$. It follows that $\{ \epsilon_n (\lambda) \}$
are even functions of $\lambda$,
$$ \epsilon_n (-\lambda) = \epsilon_n (\lambda) \,. \eqlabel{even} $$
Having solved
for these functions, one can then use the constraint equations
\eqref{constraint} to determine the equilibrium densities
$\{ \rho_n^{eq} (\lambda) \}$ and $\{ \tilde \rho_n^{eq} (\lambda) \}$.

The free energy per site at equilibrium is given by
$${F \over N} = -T \sum_{n=1}^\infty \int_{-\infty}^\infty d\lambda\
a_n (\lambda) \ln \left( 1 + e^{-\epsilon_{n}(\lambda)/T} \right) - {1\over 2}
H
\,. \eqlabel{free1} $$
This result is obtained by using the constraint equations \eqref{constraint}
to eliminate $\tilde \rho_n$ from the expression for $F/N$, and then noting
that the coefficient of $\rho_n$ vanishes by virtue of the TBA equations.
Alternatively, by eliminating $\tilde \rho_n$, one can show that
$${F\over N} = e_0
- T \int_{-\infty}^\infty d\lambda\ s(\lambda)
\ln \left[ 1 + e^{\epsilon_{1}(\lambda)/T} \right] \,, \eqlabel{free2} $$
where $e_0$ is given by \eqref{ezero}. This expression for the equilibrium
free energy is particularly useful, since it depends on only {\it one}
of the $\epsilon_n(\lambda)$'s.

\vskip 0.4truein
\noindent
{\bf \chapnum . Thermodynamics for small $T$ and $H$}
\vskip 0.2truein

In this section, we shall solve the TBA equations \eqref{TBA} for
$\epsilon_1(\lambda)$ for small values of $T$ and $H$, and evaluate the
expression \eqref{free2} for the free energy.

As we have seen, the TBA equations are an infinite set of coupled equations.
A significant simplification occurs for $T \rightarrow 0$: namely, the equation
for $\epsilon_1(\lambda)$ becomes decoupled from the other equations.
In order to see this, we first cast the TBA equations in the form given
by Takahashi${}^{\refref{takahashi1}}$:
$$\eqalignno{
\epsilon_1 & = -\pi a_1 + H +
a_2 * T \ln \left( 1 + e^{-\epsilon_1/T} \right) \cr
& \qquad\qquad
+ \left( a_0 + a_2 \right) * \sum_{m=1}^\infty a_m * T\ln \left( 1 +
e^{-\epsilon_{m+1}/T} \right)  \,, \eqalignlabel{first} \cr
\epsilon_n & = H +
a_1 * T \ln \left( 1 + e^{\epsilon_{n-1}/T} \right) +
a_2 * T \ln \left( 1 + e^{-\epsilon_n/T} \right) \cr
& \qquad\qquad
+ \left( a_0 + a_2 \right) * \sum_{m=1}^\infty a_m * T\ln \left( 1 +
e^{-\epsilon_{m+n}/T} \right) \,, \qquad n \ge 2 \,, \eqalignnum \cr} $$
where $a_0 (\lambda) = \delta (\lambda)$.
To obtain \eq\eqref{first}, we start by writing the $n=1$ TBA equation
using the explicit form of $A_{1m}$ as given in \eqref{Anm}. We then
rearrange the terms and use the identities
$$ a_n * a_m = a_{n+m} \,. \eqnum $$
To obtain the equation for $\epsilon_2$, we use the additional identity
$$ a_n * H = H  \eqnum $$
to write the driving term $-\pi a_2 + 2H$ of the $n=2$ TBA equation as
$a_1*\left( -\pi a_1 + H\right) + H$, and we then use \eq\eqref{first}
to eliminate the terms in parenthesis. The other equations are obtained by
following the same strategy.

{}From this form of the equations, we conclude that
$$ \epsilon_n (\lambda) \ge 0 \quad {\hbox{ for  }} n \ge 2 \,.
\eqlabel{positivity} $$
This positivity property will play an important role in the arguments
that follow.\footnote*{An alternative way of demonstrating this
positivity property relies on the ``inverted'' TBA equations,
which are obtained by forming the convolution of the TBA equations with
$A_{nm}^{-1}$. However, these equations must be supplemented with
a condition on $\epsilon_n$ for $n \rightarrow \infty$.}

Consider now \eq\eqref{first}. The positivity property \eqref{positivity}
implies that $\exp (- \epsilon_{m+1}/T)$ goes to zero
exponentially as $T \rightarrow 0$ for $m \ge 1$ and can therefore
be neglected. This is the reason for the decoupling of equations
mentioned above.
On the other hand, $\epsilon_1(\lambda)$ can have either sign.
Defining $\varepsilon(\lambda)$ to be the $T \rightarrow 0$ limit of
$\epsilon_1(\lambda)$, we see that
$$ \lim_{T \rightarrow 0}\ T \ln \left( 1 + e^{\pm\epsilon_1/T} \right)
= \pm \varepsilon^\pm \,, \eqlabel{limit} $$
where we use the standard notation
$$\varepsilon^- \equiv {1\over 2} \left( \varepsilon - |\varepsilon| \right)
\,, \qquad\qquad
\varepsilon^+ \equiv \varepsilon - \varepsilon^-  \,. \eqnum $$
It follows from \eqref{first} that
$$\varepsilon = -\pi a_1 + H - a_2 * \varepsilon^- \,. \eqlabel{almost} $$

It will be convenient for subsequent analysis to work with the following
equivalent equation, which does not involve $\varepsilon^-$:
$$\varepsilon = {1\over 2} H -\pi s + h * \varepsilon^+
\,, \eqlabel{transformed}  $$
where $s(\lambda)$ is defined in \eqref{s}, and $h(\lambda)$ is defined by
$$ h = s * a_1 \,. \eqlabel{h} $$
In order to obtain \eqref{transformed}, we substitute
$\varepsilon^- = \varepsilon - \varepsilon^+$ into \eqref{almost}; and
we then solve for $\varepsilon$ in terms of $\varepsilon^+$ with the
help of Fourier transforms.
We observe for future reference that by a similar procedure,
\eqref{first} can be recast (after neglecting the infinite sum,
but before taking the $T \rightarrow 0$ limit of $\epsilon_1$) as follows:
$$
\epsilon_1 = {1\over 2} H - \pi s + h * T \ln \left( 1 + e^{\epsilon_1/T}
\right)  \,. \eqlabel{transformed2}  $$
Evidently, the $T \rightarrow 0$ limit of this equation gives
\eqref{transformed}.

We have obtained the linear integral equation \eqref{transformed}
for $\varepsilon$. Of central importance
in obtaining this result was the positivity property \eqref{positivity}.
We shall now show that \eq\eqref{transformed} can be solved, and that
it has a unique solution. We shall also establish certain properties
of the solution, which will turn out to be crucial for formulating the linear
integral equation obeyed by the leading correction to this lowest-order
result.

We first briefly consider the case $H=0$, which corresponds to the ground
(vacuum) state. Since $s(\lambda)$  is positive for all $\lambda$,
\eq\eqref{transformed} has the solution
$$\varepsilon(\lambda) = -\pi s(\lambda) = -{\pi\over 2\ch \pi \lambda}
\,. \eqlabel{vacuum}  $$
{}From the definition \eqref{epsilon}, the constraint equation
\eqref{constraint}, and the positivity property \eqref{positivity},
it follows that the ground-state densities of holes and particles are
precisely those anticipated in the previous section. (See Eqs.
\eqref{noholes} and \eqref{rhozero}, respectively.)

We turn now to the case $H \ne 0$. We observe, following Yang and Yang
and Takahashi, that \eq\eqref{transformed} can be solved by iteration.
Indeed, set
$$\eqalignno{
\varepsilon_{(0)}(\lambda) &= {1\over 2} H -\pi s(\lambda) \,,
\eqalignlabel{iteratezero} \cr
\varepsilon_{(n)}(\lambda) &= {1\over 2} H -\pi s(\lambda)
+ \left( h * \varepsilon^+_{(n-1)} \right) (\lambda) \,, \qquad n \ge 1
\,. \eqalignlabel{iteraten} \cr} $$
If the sequence $\{ \varepsilon_{(0)}(\lambda) \,,
\varepsilon_{(1)}(\lambda) \,, \cdots \}$ has a limit
$$ \lim_{n \rightarrow \infty} \varepsilon_{(n)}(\lambda) =
\varepsilon(\lambda) \,, \eqnum $$
then this limit satisfies \eq\eqref{transformed}. In order to show that
this sequence is convergent, we prove that it is monotonically increasing
(i.e., $\varepsilon_{(n)}(\lambda) > \varepsilon_{(n-1)}(\lambda)$) and
that it is bounded from above. To demonstrate that the sequence is
monotonically increasing, we proceed by induction, using the fact that
$h(\lambda)$ is positive for all $\lambda$, and that
$$\varepsilon_{(n)} = \varepsilon_{(n-1)}
+ h * \left( \varepsilon^+_{(n-1)} - \varepsilon^+_{(n-2)} \right)
\,, \qquad n \ge 2 \,. \eqnum $$
To show that the sequence is bounded from above, we cast the relation
\eqref{iteraten} in the form
$$\varepsilon_{(n)} = \varepsilon_{(0)} + h * \varepsilon^+_{(0)} + \cdots
+ h * \cdots * h * \varepsilon^+_{(0)}
\,, \eqnum $$
where in the last term the factor $h$ appears $n$ times. From
\eqref{iteratezero}, we see that $\varepsilon_{(0)}(\lambda) < H/2$.
It follows that
$$ h * \cdots * h * \varepsilon^+_{(0)}(\lambda) <
\left( {1\over 2} \right)^n {H\over 2} \,, \eqnum $$
and therefore
$$ \varepsilon_{(n)}(\lambda) < {2^{n+1} - 1 \over 2^{n+1}} H \,. \eqnum $$
We conclude that
$$\varepsilon_{(n)}(\lambda) < H    \eqnum $$
for any $n$, and so the sequence is bounded from above. This completes the
proof that the iteration procedure \eqref{iteratezero}, \eqref{iteraten}
gives a solution of \eq\eqref{transformed}.

{}From this iterative solution, we can understand the general behavior of
$\varepsilon(\lambda)$ as a function of $\lambda$. For
$\lambda > 0$, this function is monotonically increasing.
Indeed, following again Refs. \refref{yangyang1} and \refref{takahashi1},
we prove the monotonicity of $\varepsilon_{(n)}(\lambda)$ by
induction. From \eqref{iteratezero}, we determine that
$\varepsilon_{(0)}'(\lambda) > 0$ for $\lambda >0$.
Differentiating \eqref{iteraten} with respect to $\lambda$, we obtain
$$\eqalignno{
\varepsilon_{(n)}'(\lambda) &= \varepsilon_{(0)}'(\lambda)
+ \int_{-\infty}^{\infty} h(\lambda - \zeta)\
\varepsilon_{(n-1)}^{+ \prime}(\zeta)\ d\zeta \eqalignnum \cr
                            &= \varepsilon_{(0)}'(\lambda)
+ \int_0^{\infty} \left[ h(\lambda - \zeta)
- h(\lambda + \zeta) \right]\
\varepsilon_{(n-1)}^{+ \prime}(\zeta)\ d\zeta \,.
\eqalignlabel{monoton} \cr} $$
Since the function $h(\lambda)$ is even and monotonically decreasing
for $\lambda > 0$, the expression inside the brackets in
\eq\eqref{monoton} is positive. Therefore, if
$\varepsilon_{(n-1)}'(\lambda) > 0$ for $\lambda > 0$, then
$\varepsilon_{(n)}'(\lambda) > 0$ for $\lambda > 0$. This concludes
the proof that $\varepsilon_{(n)}(\lambda)$ (and hence
$\varepsilon(\lambda)$) is a monotonic increasing function of $\lambda$,
for $\lambda > 0$.

On the other hand, we have seen that $\varepsilon (\lambda)$
is bounded from above. Therefore, the function $\varepsilon (\lambda)$
has the following general behavior:

\vskip 6pt
\centerline{figure available on request}
\vskip 6pt

\noindent
In particular, $\varepsilon (\lambda)$ has only one zero for $\lambda > 0$,
$$ \varepsilon(\alpha) =  0  \,. \eqnum $$
Hence, $\varepsilon(\lambda) < 0$ for $\lambda$ in the interval
$\left( 0 \,, \alpha \right)$, and
$\varepsilon(\lambda) > 0$ for $\lambda$ in the interval
$\left( \alpha \,, \infty \right)$.

Observe that \eq\eqref{transformed} is temperature-independent. Using the
same approximations in the expression \eqref{free2} for the free energy this
too
will be temperature-independent. As emphasized by Johnson and
McCoy${}^{\refref{johnson/mccoy}}$, to find the leading order temperature
dependence one must compute the leading correction to the
solution $\epsilon_1 = \varepsilon$ of the linearized equation
\eqref{transformed}. In order to obtain this correction, we make the
substitution
$$\epsilon_1 (\lambda) = \varepsilon(\lambda)
+ \eta (\lambda) \eqlabel{expansion} $$
in \eqref{transformed2} and expand to leading order in $\eta$.
Since $\varepsilon$ is a solution of the linearized equation
\eqref{transformed}, we shall find an inhomogeneous term in the resulting
equation for $\eta$. Indeed, we have that
$$\eta = h * \left\{ T \ln \left[ 1 +
e^{\left( \varepsilon + \eta \right)/T} \right]
- \varepsilon^+ \right\} \,. \eqlabel{indeed}  $$
Because $\varepsilon(\lambda)$ is
an even function of $\lambda$ with a single zero, at $\lambda = \alpha$,
for positive $\lambda$, we see that
$$\eqalignno{
\eta (\lambda) &= \left( \int_{-\infty}^{-\alpha} +
\int_\alpha^\infty \right) d\lambda'\ h(\lambda - \lambda')
\Big\{ T \ln \left[ 1 +
e^{\left( \varepsilon(\lambda') + \eta(\lambda') \right)/T} \right]
- \varepsilon(\lambda') \Big\} \cr
& \quad + \int_{-\alpha}^\alpha d\lambda'\ h(\lambda - \lambda')\
T \ln \left[ 1 + e^{\left( \varepsilon(\lambda') + \eta(\lambda')
\right)/T} \right] \cr
& \approx \left( \int_{-\infty}^{-\alpha} +
\int_\alpha^\infty \right) d\lambda'\ h(\lambda - \lambda')\
\eta(\lambda') + E_h(\lambda) \eqalignnum \cr}
$$
where the inhomogeneous term $E_h(\lambda)$ is given by
$$ E_h = h * T \ln \left( 1 + e^{-|\varepsilon|/T} \right)
\,. \eqnum $$
For $T \rightarrow 0$, the major contribution to the integral comes from
the regions near the zeros of $\varepsilon$, so we expand
$\varepsilon (\lambda)$ about $\lambda = \alpha$,
$$ \varepsilon (\lambda)  = t \left( \lambda - \alpha \right)
+ O \left( (\lambda - \alpha)^2 \right) \,, \qquad\qquad
t \equiv {d \varepsilon \over d\lambda}
\Big\vert_{\lambda = \alpha} \,. \eqlabel{tdefine} $$
One then finds that the leading $T$- dependence of $E_h$ is
$$\eqalignno{
E_h(\lambda) &= {2 T^2\over t} \left[ h(\lambda - \alpha) +
h(\lambda + \alpha) \right] \int_0^\infty du\ \ln \left( 1 + e^{-u}
\right) \cr
             &= {\pi^2 T^2 \over 6 t} \left[ h(\lambda - \alpha) +
h(\lambda + \alpha) \right] \,. \eqalignnum \cr} $$
Thus, we obtain the following linear, but nonlocal, equation for
$\eta$,\footnote*{While Johnson and McCoy obtain this equation from
the ``inverted'' TBA equations, we obtain it directly from the
TBA equations.}
$$\eta(\lambda) =  \left( \int_{-\infty}^{-\alpha} +
\int_\alpha^\infty \right) d\lambda'\ h(\lambda - \lambda')\ \eta(\lambda')
+ {\pi^2 T^2 \over 6 t} \left[ h(\lambda - \alpha) +
h(\lambda + \alpha) \right] \,. \eqlabel{correction} $$
This equation for $\eta$ and that of \eqref{transformed} for
$\varepsilon$ complete our results for the $T$ - expansion of
$\epsilon_1$. We turn now to the expansion for small values of $H$,
by which from Eqs. \eqref{transformed} and \eqref{correction} we shall generate
a pair of integral equations of the Wiener-Hopf type. (For a detailed
exposition of Wiener-Hopf equations, see Ref. \refref{krein}.)

For $H \rightarrow 0$, we see from the solution of the zeroth-order equation
\eqref{iteratezero} that the function $\varepsilon(\lambda)$ has
a zero for $\lambda = \alpha = O \left( \ln H \right)$. Assuming an expansion
in powers of $(\ln H)^{-1}$, we conclude that
$$\alpha = -{1\over \pi} \left[
\ln \left( {1\over 2\pi} H \right) + \ln \kappa +
O \left( {1\over \ln H} \right) \right] \,, \eqlabel{kappa} $$
where the constant $\kappa$ (which is independent of $H$) has still to be
determined.

It will prove convenient to work with the functions
$$S(\lambda) = \left\{ \matrix{e^{\pi \alpha}\
\kappa\ \varepsilon (\lambda + \alpha)  & \lambda > 0 \cr
0 & \lambda < 0 \cr} \right. \,, \eqlabel{funcS} $$
and
$$T(\lambda) = \left\{ \matrix{ {6 e^{-\pi \alpha}
\over \pi^2 T^2 \kappa}
\eta (\lambda + \alpha)  & \lambda > 0 \cr
0 & \lambda < 0 \cr} \right. \,, \eqlabel{funcT} $$
instead of the functions $\varepsilon(\lambda)$ and
$\eta(\lambda)$.
The factors $e^{\pi\alpha}$ and
$e^{-\pi \alpha} /T^2$ in \eqref{funcS}
and \eqref{funcT}, respectively, are chosen such that the driving terms
in the equations for $S(\lambda)$ and $T(\lambda)$ have a
nonvanishing limit as $T \rightarrow 0$ and $H \rightarrow 0$.

We return now to \eq\eqref{transformed}. Following Refs. \refref{yangyang2}
and \refref{johnson/mccoy}, we write the limits of
integration explicitly, keeping in mind that $\varepsilon(\lambda) > 0$
for $-\infty < \lambda <  -\alpha$ and for
$\alpha < \lambda < \infty $; and we shift the integration
variables so that they run from $0$ to $\infty$. Since
$h\left( \lambda + 2\alpha \right)$ vanishes as $H \rightarrow 0$
($\alpha \rightarrow \infty$) for finite $\lambda$, we see that
the $H \rightarrow 0$ limit of the equation for $S(\lambda)$ is
$$S(\lambda) = \pi \left( 1 - \kappa e^{-\pi\lambda} \right)
+ \int_0^\infty d\lambda'\ h(\lambda - \lambda')\ S(\lambda')  \,,
\qquad \lambda \ge 0 \,.
\eqnum $$
Similarly, from \eqref{correction}, we see that the $H \rightarrow 0$ limit
of the equation for $T(\lambda)$ is
$$T(\lambda) = {h(\lambda)\over S'(0)} + \int_0^\infty d\lambda'\
h(\lambda - \lambda')\ T(\lambda')  \,, \qquad \lambda \ge 0 \,,
\eqnum $$
where
$$ S'(0) \equiv
{d\over d\lambda}S(\lambda)\Big\vert_{\lambda = 0+}
= e^{\pi \alpha} \kappa\ t \,. \eqlabel{sprime} $$

These equations can be written in the standard Wiener-Hopf form
$$\eqalignno{
S(\lambda) &= f_S(\lambda) + b_S(\lambda)
+ \int_{-\infty}^\infty d\lambda'\
h(\lambda - \lambda')\ S(\lambda')  \,, \cr
T(\lambda) &= f_T(\lambda) + b_T(\lambda)
+ \int_{-\infty}^\infty d\lambda'\
h(\lambda - \lambda')\ T(\lambda')   \,, \cr
&\qquad\qquad\qquad\qquad\qquad\qquad\qquad\qquad\qquad
 -\infty < \lambda < \infty \,,
\eqalignnum \cr} $$
where
$$f_S(\lambda) = \left\{ \matrix{
\pi \left( 1 - \kappa e^{-\pi\lambda} \right)
& \lambda > 0 \cr
0 & \lambda < 0 \cr} \right. \,, \eqlabel{fS} $$
$$b_S(\lambda) = \left\{ \matrix{
0 & \lambda > 0 \cr
- \int_{-\infty}^\infty d\lambda'\
h(\lambda - \lambda')\ S(\lambda')
& \lambda < 0 \cr} \right. \,, \eqlabel{bS} $$
and similarly
$$f_T(\lambda) = \left\{ \matrix{
{h(\lambda)\over S'(0)} & \lambda > 0 \cr
0 & \lambda < 0 \cr} \right. \,, \eqlabel{fT} $$
$$b_T(\lambda) = \left\{ \matrix{
0 & \lambda > 0 \cr
- \int_{-\infty}^\infty d\lambda'\
h(\lambda - \lambda')\ T(\lambda')
& \lambda < 0 \cr} \right. \,. \eqlabel{bT} $$

We shall solve these equations by Fourier transform.
Since $S(\lambda)$ and $T(\lambda)$ vanish for $\lambda < 0$,
the Fourier transforms $\hat S(\omega)$ and $\hat T(\omega)$ are
analytic in the upper-half-plane Im $\omega \ge 0$, which we
denote by $\Pi_+$. Observe that
$$S(0) = {2\over 2\pi} \int_{-\infty}^\infty d\omega\ \hat S(\omega) \,,
\eqnum $$
where the factor 2 in the numerator is due to the discontinuity of $S(\lambda)$
at $\lambda=0$. Evidently, the contour integral of $\hat S(\omega)$ along
a closed contour in $\Pi_+$ is zero. By explicitly evaluating the contribution
of the semi-circle at infinity, we obtain
$$\int_{-\infty}^\infty d\omega\ \hat S(\omega)
= - \int_{semi-circle} d\omega\ \hat S(\omega)= -i \pi
\lim_{|\omega| \rightarrow \infty}\ \omega \hat S(\omega) \,, \eqnum $$
where the limit is taken in $\Pi_+$. On the other hand,
$S(0) = 0$, as follows from $\varepsilon(\alpha) = 0$. Therefore,
$$ S(0) =
-i \lim_{|\omega| \rightarrow \infty}\ \omega \hat S(\omega)
= 0  \,. \eqlabel{bc} $$
Similarly,
$$ S'(0) =
{d\over d\lambda}S(\lambda)\Big\vert_{\lambda = 0+}
= - \lim_{|\omega| \rightarrow \infty}\ \omega^2 \hat S(\omega)
\,. \eqlabel{second} $$
In deriving this result, we must again include a factor 2 due to the
discontinuity (here, of $S'(\lambda)$) at $\lambda = 0$.

The Wiener-Hopf equations in Fourier space are
$$\eqalignno{
\hat S(\omega) &= \hat f_S(\omega) + \hat b_S(\omega)
+ \hat h(\omega)\ \hat S(\omega)  \,, \eqalignlabel{WHS} \cr
\hat T(\omega) &= \hat f_T(\omega) + \hat b_T(\omega)
+ \hat h(\omega)\ \hat T(\omega)  \,. \eqalignlabel{WHT} \cr} $$

Since $\hat h(-\omega) = \hat h(\omega)$ and $1 - \hat h(\omega) \ne 0$
for $-\infty < \omega < \infty$, there exists${}^{\refref{krein}}$ a unique
factorization
$$ \left( 1 - \hat h(\omega) \right)^{-1} = G_+ (\omega)\ G_- (\omega)
\,, \qquad -\infty < \omega < \infty \,,
\eqlabel{factorization} $$
where $G_+ (\omega)$ and $G_- (\omega)$ are functions that are analytic
and nonvanishing in the half-planes $\Pi_+$ and $\Pi_-$, respectively,
and are normalized by the condition
$$ G_+(\infty) = G_-(\infty) = 1 \,. \eqnum $$
Moreover, the fact $h(-\lambda) = h(\lambda)$ implies that (for
$\omega$ in $\Pi_-$)
$$G_-(\omega) = G_+(-\omega) \,. \eqlabel{Gproperty} $$
As we shall see, explicit expressions for $G_+$ and $G_-$ are not needed
to compute the free energy to the order in which we work. A similar
phenomenon occurs in the Wiener-Hopf calculations of Yang and
Yang${}^{\refref{yangyang2}}$.

These properties will be used to solve the Wiener-Hopf equations
\eqref{WHS} and \eqref{WHT}, to which we now turn.
Using the factorization \eqref{factorization}, the equation
\eqref{WHS} for $\hat S$ can be rewritten as
$$G_+^{-1}\ \hat S = G_- \left( \hat f_S + \hat b_S \right) \,.
\eqlabel{interS} $$
Observe that the left hand side is analytic and bounded in $\Pi_+$,
whereas $G_- \hat b_S$ is analytic and bounded in $\Pi_-$. The term
$G_- \hat f_S$ has a decomposition as
$$G_- \hat f_S = P_- \left( G_- \hat f_S \right) +
P_+ \left( G_- \hat f_S \right) \,, \eqlabel{decomposition} $$
where $P_\pm \left( G_- \hat f_S \right)$ is analytic in $\Pi_\pm$.
This decomposition is uniquely specified by the requirement
$P_\pm \left( G_- \hat f_S \right) \rightarrow 0$ for $\omega \rightarrow
\infty$ in $\Pi_\pm$.
Taking the $P_+$ projection of \eqref{interS}, we have that
$$\hat S = G_+\ P_+ \left( G_- \hat f_S \right) \,. \eqlabel{formalS} $$
{}From \eqref{fS} we compute that
$$\hat f_S (\omega) = \pi i\left(
{1\over \omega + i\epsilon} - {\kappa\over \omega + \pi i}
\right) \,, \eqnum $$
where one is to take $\epsilon \rightarrow 0$ at the end.
The decomposition \eqref{decomposition} of $G_- \hat f_S$ is then found
by subtracting the residues of $\hat f_S$, i.e.,
$$\eqalignno{
G_-(\omega) \hat f_S (\omega) =& \pi i \left\{
{1\over \omega + i\epsilon}
\left( G_-(\omega) - G_-(-i\epsilon) \right)
-{\kappa\over \omega + \pi i}
\left( G_-(\omega) - G_-(-\pi i) \right) \right\} \cr
&+ \pi i\left\{
{1\over \omega + i\epsilon} G_-(-i\epsilon)
-{\kappa\over \omega + \pi i} G_-(-\pi i) \right\} \,. \eqalignnum \cr}$$
Hence,
$$\hat S(\omega) = {\pi i \over \omega + i\epsilon}
G_+(\omega) G_-(0)
- {\pi i \kappa\over \omega + \pi i}
G_+(\omega) G_-(-\pi i) \,. \eqnum $$
The boundary condition \eqref{bc} implies that the parameter $\kappa$
is given by
$$ \kappa = G_-(-\pi i)^{-1}\  G_-(0) \,.
\eqlabel{explicitkappa} $$
Using this result we conclude that $\hat S(\omega)$ is given by
$$\hat S(\omega) = \pi i
\left( {1\over \omega + i\epsilon} - {1\over \omega + \pi i} \right)
G_+(\omega) G_-(0) \,. \eqlabel{Ssolution} $$
{}From \eq\eqref{second} and the fact that
$G_+ (\omega) \rightarrow 1$ as $|\omega| \rightarrow \infty$ in $\Pi_+$,
we find that
$$ S'(0)
= \pi^2  G_-(0) \,. \eqlabel{derivative} $$

We turn now to the equation \eqref{WHT} for $\hat T(\omega)$.
Proceeding as before, we use the factorization \eqref{factorization}
to arrive at the formal solution
$$\hat T = G_+\ P_+ \left( G_- \hat f_T \right)  \eqlabel{formalT} $$
(cf. \eq\eqref{formalS} ). The explicit calculation of $\hat f_T$ is
difficult, but can be avoided by the following trick. Consider the
functions $f(\lambda)$ defined as
$$f(\lambda) =  {h(\lambda)\over S'(0)} \,, \eqnum $$
where $\lambda$ ranges over the entire real line. From Eq. \eqref{fT}, it is
evident that
$$f_T(\lambda) = f_+(\lambda) \,, \eqnum $$
where
$$ f_+(\lambda) \equiv \left\{ \matrix{ f(\lambda) & \lambda > 0 \cr
                                        0          & \lambda < 0 \cr} \right.
\,, \qquad\qquad
   f_-(\lambda) \equiv \left\{ \matrix{   0          & \lambda > 0 \cr
                                          f(\lambda) & \lambda < 0 \cr} \right.
\,, \qquad\qquad f = f_+ + f_- \,.
\eqnum $$
The Fourier transform of $f(\lambda)$ is readily computed:
$$\hat f(\omega) = {\hat h(\omega) \over S'(0)} \,. \eqnum $$
{}From the factorization \eqref{factorization}, it follows that
$$\hat f_+ + \hat f_- = \left( 1 - G_-^{-1} G_+^{-1} \right)
{1\over S'(0)} \,. \eqnum $$
After multiplying both sides of this equation by $G_-$, we see that
the $P_+$ projection of $G_- \hat f_+$ is given by
$$P_+ \left( G_- \hat f_+ \right) = \alpha - {G_+^{-1} \over S'(0)}
\,, \eqlabel{Q+} $$
where $\alpha$ is a constant. Requiring the right hand side to vanish
for $|\omega| \rightarrow \infty$ in $\Pi_+$ determines this constant to
be
$$\alpha = {1\over S'(0)} \,. \eqnum $$
We conclude from \eqref{formalT} that $\hat T(\omega)$ is given by
$$\hat T (\omega) = \left( G_+(\omega) - 1 \right) {1\over S'(0)}
 \,, \eqlabel{Tsolution} $$
where $S'(0)$ is given by \eqref{derivative}.

To summarize: we have made the expansion \eqref{expansion}
of $\epsilon_1(\lambda)$, and we have changed in Eqs. \eqref{funcS},
\eqref{funcT} from the variables
$\varepsilon(\lambda)$, $\eta(\lambda)$ to the variables
$S(\lambda)$, $T(\lambda)$, respectively. Using Wiener-Hopf
methods, we have determined in Eqs. \eqref{Ssolution}, \eqref{Tsolution} the
corresponding Fourier transforms $\hat S(\omega)$, $\hat T(\omega)$
in the limits $T \rightarrow 0$ and $H \rightarrow 0$. These expressions
involve $G_+(\omega)$ and $G_-(\omega)$, which appear in the factorization
\eqref{factorization}. We shall now use these results
to calculate the free energy.

Substituting the expansion \eqref{expansion} of $\epsilon_1 (\lambda)$ into
the expression \eqref{free2} for the free energy (keeping in mind the
discussion immediately following \eqref{expansion}), we obtain
$${F\over N} = e_0
-{\pi^2 T^2 s(\alpha) \over 3t}
- 2 \int_\alpha^\infty  d\lambda\ s(\lambda)
\left[  \varepsilon(\lambda) +  \eta(\lambda) \right]
\,, \eqlabel{freeint} $$
where the ground state energy per site $e_0$ is given by
\eqref{ezero}.

We aim for a double expansion of $F/N$ to quadratic order in both $T$ and $H$.
To this end, we rewrite \eqref{freeint} in terms of $S(\lambda)$ and
$T(\lambda)$.
Consider first the $T^2$ term. For $H \rightarrow 0$, we can make
the approximation
$$ s(\alpha) = e^{-\pi \alpha}
\,, \eqnum $$
and hence (remembering \eqref{sprime}) this term becomes
$$ -{\pi^2 T^2 \kappa \over 3 S'(0)} \,. \eqnum $$
We next manipulate the integral in the last term as follows:
$$
\int_\alpha^\infty d\lambda\ s(\lambda)
\left[  \varepsilon(\lambda) +  \eta(\lambda) \right]
= \int_0^\infty  d\lambda\ s(\lambda + \alpha)
\left[ {e^{-\pi \alpha}\over \kappa} S(\lambda)
+  {\pi^2 T^2 \kappa
\over 6 e^{-\pi \alpha}} T(\lambda) \right] \,. \eqnum $$
Taking the $\alpha \rightarrow \infty$ limit of this expression,
we arrive at the following expression for $F/N$,
$${F\over N} = e_0 - {H^2 \kappa \over 2\pi^2}
\int_0^\infty  d\lambda\ e^{-\pi \lambda}  S(\lambda)
- {\pi^2 T^2 \kappa \over 3}  \left[
\int_0^\infty  d\lambda\ e^{-\pi \lambda} T(\lambda) +
{1\over S'(0)} \right] \,. \eqnum  $$
Writing the functions $S(\lambda)$ and $T(\lambda)$
as the Fourier transforms of $\hat S(\omega)$ and $\hat T(\omega)$,
respectively, and then performing both the $\lambda$ and $\omega$
integrals, we obtain
$${F\over N} = e_0
-{H^2 \over 2\pi^2} A  - {\pi^2 T^2 \over 3} B \,, \eqnum $$
with
$$ A = \kappa \hat S(\pi i) \,, \qquad\qquad
   B = \kappa \left[ \hat T(\pi i) + {1\over S'(0)} \right]
\,. \eqnum  $$

The parameter $\kappa$ is given in \eq\eqref{explicitkappa};
and using the property \eqref{Gproperty}, this parameter can
be reexpressed as
$$\kappa = G_+(0) G_+(\pi i)^{-1}  \,. \eqnum $$

We consider now $A$. Evaluating $\hat S(\pi i)$ using the expression
\eqref{Ssolution}, we obtain
$$ A = {1\over 2} G_+(0) G_-(0) \,. \eqnum $$
Recalling the factorization equation \eqref{factorization} and the
fact that
$$ \hat h(\omega) = \hat s(\omega)\ \hat a_1(\omega) \,, \eqnum $$
we conclude that
$$ A = {1\over 2} \left( 1 - \hat h(0) \right)^{-1}  = 1 \,. \eqnum $$

There remains to compute $B$. Evaluating $\hat T(\pi i)$ using the
expression \eqref{Tsolution}, we obtain (after a crucial cancelation)
$$ B = {G_+(0) \over S'(0)} \,. \eqnum $$
Recalling the expression \eqref{derivative} for $S'(0)$ and using again the
property \eqref{Gproperty}, we see that
$$ B = {1 \over \pi^2} \,.
\eqnum $$

The expression for the free energy per site is therefore
$${F\over N} = e_0 - {1\over 2\pi^2} H^2 - {1\over 3}T^2 \,. \eqnum $$
As foreseen above, this result was obtained without using
explicit expressions for $G_+$ and $G_-$.
It follows that the magnetic susceptibility and specific heat, to lowest
order, are given by
$$\eqalignno{
\chi &= - {\partial^2\over \partial H^2} \left( {F\over N} \right)
\Big\vert_T = {1\over \pi^2} \,, \cr
C_H &= -T {\partial^2\over \partial T^2} \left( {F\over N} \right)
\Big\vert_H = {2\over 3}T \,.  \eqalignnum \cr} $$

For a critical chain, the free energy per site is given
by${}^{\refref{critical}}$
$${F\over N} = e_0 - {\pi c\over 6 v_s} T^2 + \cdots \,, \eqnum $$
where $c$ is the central charge and $v_s$ is the velocity of sound.
For the antiferromagnetic Heisenberg Hamiltonian \eqref{xxx},
$v_s = \pi/2$ (see, e.g., Ref. \refref{faddeev1}), and hence $c=1$.

\vskip 0.4truein
\noindent
{\bf \chapnum . Acknowledgments}
\vskip 0.2truein

It is a pleasure to acknowledge E. Melzer and A.M. Tsvelik for various
discussions. We thank the Editors for inviting us to the Workshop,
and for giving us the opportunity to present this review. Part of this work
was performed at the Aspen Center for Physics. This work was supported in part
by the National Science Foundation under Grants No. PHY-90 07517 and
PHY-92 09978.

\vskip 0.4truein
\noindent
{\bf \chapnum . References}
\vskip 0.2truein

\reflabel{qsymmetry}
V.G. Drinfel'd, J. Sov. Math. {\it 41} (1988) 898;
M. Jimbo, Int'l J. Mod. Phys. {\it A4} (1989) 3759;
L.D. Faddeev, N. Yu. Reshetikhin and L.A. Takhtajan, Algebraic Analysis,
{\it 1} (1988) 129;  Leningrad Math. J. {\it 1} (1990) 193.

\reflabel{alcaraz}
F.C. Alcaraz, M.N. Barber, M.T. Batchelor, R.J. Baxter and G.R.W. Quispel,
J. Phys. {\it A20} (1987) 6397.

\reflabel{sklyanin}
E.K. Sklyanin, J. Phys. {\it A21} (1988) 2375.

\reflabel{pasquier}
V. Pasquier and H. Saleur, Nucl. Phys. {\it B330} (1990) 523.

\reflabel{kulish/sklyanin(1)}
P.P. Kulish and E.K. Sklyanin, J. Phys. {\it A24} (1991) L435.

\reflabel{spin1}
L. Mezincescu, R.I. Nepomechie and V. Rittenberg, Phys. Lett. {\it A147}
(1990) 70.

\reflabel{general}
L. Mezincescu and R.I. Nepomechie, Int'l J. Mod. Phys. {\it A6} (1991) 5231;
Addendum, {\it A7} (1992) 5657;
Mod. Phys. Lett. {\it A6} (1991) 2497; Nucl. Phys. {\it B372} (1992) 597.
This work is reviewed in our contribution to the Louis Witten
Festschrift (World Scientific), in press.

\reflabel{batchelor}
M.T. Batchelor and A. Kuniba, J. Phys. {\it A24} (1991) 2599.

\reflabel{yangyang1}
C.N. Yang and C.P. Yang, J. Math. Phys. {\it 10} (1969) 1115.

\reflabel{takahashi1}
M. Takahashi, Prog. Theor. Phys. {\it 46} (1971) 401.

\reflabel{gaudin}
M. Gaudin, Phys. Rev. Lett. {\it 26} (1971) 1301;
{\it La fonction d'onde de Bethe} (Masson, 1983).

\reflabel{johnson/mccoy}
J.D. Johnson and B.M. McCoy, Phys. Rev. {\it A6} (1972) 1613.

\reflabel{takahashi2}
M. Takahashi, Prog. Theor. Phys. {\it 50} (1973) 1519;
{\it 51} (1974) 1348.

\reflabel{filyov}
V.M. Filyov, A.M. Tsvelik and P.B. Wiegmann, Phys. Lett. {\it 81A}
(1981) 175.

\reflabel{faddeev1}
L.D. Faddeev and L.A. Takhtajan, J. Sov. Math. {\it 24} (1984) 241.

\reflabel{tsvelick/wiegmann}
A.M. Tsvelick and P.B. Wiegmann, Adv. in Phys. {\it 32} (1983) 453.

\reflabel{andrei}
N. Andrei, K. Furuya and J.H. Lowenstein, Rev. Mod. Phys. {\it 55}
(1983) 331.

\reflabel{bogoliubov}
N.M. Bogoliubov, A.G. Izergin and V.E. Korepin, in
{\it Lecture Notes in Physics} {\it 242} (Springer, 1985) 220.

\reflabel{bethe}
H. Bethe, Z. Phys. {\it 71} (1931) 205.

\reflabel{faddeev2}
L.D. Faddeev and L.A. Takhtajan, Russ. Math. Surv. {\it 34} (1979) 11.

\reflabel{reshetikhin}
N.Yu. Reshetikhin, Sov. Phys. JETP {\it 57} (1983) 691.

\reflabel{hulthen}
L. Hulthen, Arkiv. Mat. Astron. Fys. {\it 26A} (1938) 1.

\reflabel{krein}
M.G. Krein, in {\it American Mathematical Society
Translations, Series 2, Vol 22} (American Mathematical Society, 1962) 163.

\reflabel{yangyang2}
C.N. Yang and C.P. Yang, Phys. Rev. {\it 150} (1966) 327.

\reflabel{critical}
H.W.J. Bl\"ote, J.L. Cardy and M.P. Nightingale, Phys. Rev. Lett. {\bf 56}
(1986) 742; I. Affleck, Phys. Rev. Lett. {\bf 56} (1986) 746.

\end